\documentstyle[12pt]{article}
\normalbaselines

\def\iint{\int\int}
\def\vep{\varepsilon}
\def\wh{\widehat}
\def\wt{\widetilde}
\def\ore{\stackrel{\rightharpoonup}{E}}
\def\orb{\stackrel{\rightharpoonup}{B}}
\def\ole{\overleftarrow{\rightarrow}}
\def\om{\Omega}

\def\noi{\noindent}

\begin{document}
\begin{flushright}
UCVFC-DF/8-96
\end{flushright}

\vbox{\vspace{6mm}}

\begin{center}
{\bf ELECTRIC-MAGNETIC DUALITY AND THE\\ [3mm]
``LOOP REPRESENTATION'' IN ABELIAN GAUGE THEORIES}\\ [5mm]
{\it Lorenzo Leal \footnote{e-mail: lleal@dino.conicit.ve 
$\ \ $Postal address: A.P. 47399, Caracas 1041-A, Venezuela}\\ [3mm]
Departamento de F\'{\i}sica, Facultad de Ciencias\\ [3mm]
Universidad Central de Venezuela, Caracas, Venezuela.}
\end{center}

\vspace{1cm}

\begin{abstract}
\noi
Abelian Gauge Theories are quantized in a geometric representation that 
generalizes the Loop Representation and treates electric and magnetic 
operators on the same footing. The usual canonical algebra is turned into a 
topological algebra of non local operators that resembles the order-disorder 
dual algebra of 't Hooft. These dual operators provide a complete description 
of the physical phase space of the theories.
\end{abstract}

\newpage

Dual symmetry in Electromagnetism has been a source of recent developments in 
Theoretical Physics. Besides the related dual symmetry in superstrings theory 
(for a recent review and references see ref 1), the notion of duality appears 
in supersymmetric Yang-Mills  and topological theories [2-6]. There is 
also a renewed interest in the study of the analogous to electric-magnetic 
duality in conventional Yang-Mills theory [7-11].

The purpose of this letter is to study, within the geometrical language that 
the Loop Representation (L.R.) provides [12-15], the topological content 
underlying the ``electro-magnetic'' duality that ordinary Abelian gauge 
theories present. As we shall see, when these theories are quantized in a 
generalized L.R. that explicitly incorporates the dual symmetry, the 
canonical Poisson algebra results translated into a non-local algebra of 
geometric dependent operators. This algebra is topological, in the sense 
that it is not affected by continuous deformations of the spatial manifold.

To begin, let us consider free Maxwell theory in 3+1 dimensions. We shall 
work in flat space time, with $g_{\mu\nu}=diag(+,-,-,-)$. Maxwell equations
$$
\partial^\mu F_{\mu\nu}=0 \eqno (1.a)
$$
$$
\partial^\mu\  {}^*F_{\mu\nu}=0 \eqno (1.b)
$$
where $^*F_{\mu\nu}=\frac{1}{2}\vep_{\mu\nu\lambda\rho}F^{\lambda\rho}$, are 
invariant under $\ore \to-\orb ,\orb \to \ore$ (i.e.: $F\to\ ^*F$). 
Then, it is equally admisible to take for the Maxwell lagrangian, the usual 
one:
$$
{\cal{L}} = -\frac{1}{4} F_{\mu\nu}(A)F^{\mu\nu}(A) \eqno (2)
$$
with $F_{\mu\nu}(A)=\partial_\mu A_\nu -\partial_\nu A_\mu$, or a ``dual'' 
lagrangian
$$
{\wt{\cal{L}}} = -\frac{1}{4}\ {}^*F_{\mu\nu}(\wt{A})\ {}^*F^{\mu\nu}(\wt{A}) 
\eqno (3)
$$
where $^*F_{\mu\nu}(\wt{A})=\partial_\mu \wt{A}_\nu -\partial_\nu \wt{A}_\mu$.

Since both $A$ and $\wt{A}$ are one-forms obeying the same dinamical 
equations, it is often said that Maxwell theory is ``self-dual''. From the 
canonical point of view, ``self-duality'' states that the weak and strong 
coupling regimes of the theory may be exchanged. In general abelian theories 
(even in electromagnetism when $d\ne 4$), it may hapen that duality 
transformation maps one theory into another different one. For instance, as 
it is well known, the massless scalar theory is dual to the second rank gauge 
theory in $d=4$.

To some extent, both the ``direct'' and ``dual'' descriptions of Maxwell
field can be considered from the begining, by taking the first order master 
lagrangian.
$$
{\cal{L}}' = \frac{1}{2}\vep^{\mu\nu\lambda\rho}\partial_\mu A_\nu 
B_{\lambda\rho}-\frac{1}{4}B_{\lambda\rho}B^{\lambda\rho}\eqno (4)
$$
Varying $B$ and $A$ produces the equations of motion
$$
\vep^{\mu\nu\lambda\rho}\partial_\mu A_\nu =B^{\lambda\rho} \eqno (5)
$$
and 
$$
\vep^{\mu\nu\lambda\rho}\partial_\mu B_{\lambda\rho}=0 \eqno (6)
$$
respectively. Replacing $B^{\lambda\rho}$ from (5) into (4), gives back the 
usual Maxwell lagrangian (2). On the other hand, solving equation (8) yields:
$$
B_{\lambda\rho}=\partial_\lambda \wt{A}_\rho -\partial_\rho \wh{A}_\lambda 
\eqno (7)
$$
whenever the space-time manifold is ``topologically trivial'' (more 
precisely, when the $2^{nd}$ De Rham Cohomology group is trivial). Inserting 
(7) into (4) produces the ``dual'' lagrangian $\wt{\cal{L}}$.

When performing the canonical analysis from $\cal{L}'$, however, one 
encounters an undesirable feature from the point of view of the L.R. approach. 
Since the $B^2$ term breaks the evident gauge invariance (see equation (10)) 
that the first one has, one is lead to deal with second class constraints. To 
surmount this difficulty, one can add an auxiliary field $C_\mu$, and consider the 
modified lagrangian:
$$
{\cal{L}}'' = \frac{1}{2}\vep^{\mu\nu\lambda\rho}\partial_\mu A_\nu 
B_{\lambda\rho}-\frac{1}{4}(B_{\lambda\rho}+C_{\rho ,\lambda}-
C_{\lambda ,\rho})^2 \eqno (8)
$$
which is invariant under the gauge transformations
$$
\delta A_\mu = \wedge_{,\mu}\eqno (9)
$$
$$
\delta B_{\lambda\rho}=-\xi_{\rho ,\lambda}+\xi_{\lambda ,\rho}\eqno (10)
$$
$$
\delta C_\rho =\xi_\rho \eqno (11)
$$
The issue of adding an (unphysical) auxiliary field to turn the 
constraints into first class, has been used in the past [16-17] in several 
contexts. 

Once we have a full gauge-invariant theory, we can proceed to a ``L.R. 
formulation''. A lot of work is saved following Fadeev and Jackiw [18] by 
noticing that from the very structure of the ``B-F'' term:
$$
\frac{1}{2}\vep^{\mu\nu\lambda\rho}\partial_\mu A_\nu B_{\lambda\rho}=
\frac{1}{2}\vep^{ijk}(\dot A_iB_{jk}+A_0\partial_iB_{jk}+2B_{0k}\partial_i
A_j) \eqno (12)
$$
the fields $A_i$ and $\frac{1}{2}\vep^{ijk}B_{jk}$ are already canonical 
conjugates:
$$
\left [A_k(x),\frac{1}{2}\vep^{lij}B_{ij}(y)\right ] =
i\delta^l_k \delta^3(x-y)\eqno (13)
$$
Then, the Hamiltonian results to be:
$$
H=\int d^3x\left\{ \frac{1}{2}(\pi^i_C)^2+\frac{1}{4}
(F_{ij}(C)+B_{ij})^2-C_0 (\pi^i_{C,i})-\frac{1}{2}A_0\vep^{ijk}B_{jk,i}
+ B_{k0}(\pi^k_C-\vep^{ijk}A_{i,j})\right \} \eqno (14)
$$
where $\pi^i_C(x)$ are the momenta conjugate to $C_i(x)$.

The primary constraints:
$$
\pi_{A_0}\approx 0\ \ \ ,\ \ \ \pi_{C_0}\approx 0\ \ \ ,\ \ \ 
\pi_{B_{K0}}\approx 0 \eqno (15)
$$
lead to the secondary ones:
$$
\pi^i_{c,i}\approx 0 \eqno (16)
$$
$$
\vep^{ijk}B_{jk,i}\approx 0 \eqno (17)
$$
$$
\pi^k_c\approx \vep^{ijk}A_{i,j} \eqno (18)
$$
as is easily verified. Since constraint (16) is consequence of (18), the 
former may be ignored from now on. It can be checked that all the constraints 
are first class, as correspond to a ``full'' gauge invariant theory. When 
acting over physical states the Hamiltonian reduces to:
$$
H=\int d^3x\frac{1}{2}\left \{{\cal{B}}^i{\cal{B}}^i+{\cal{E}}^i{\cal{E}}^i
\right \} \eqno (19)
$$
where we have set:
$$
{\cal{B}}^k \equiv \vep^{ijk}A_{i,j} \eqno (20)
$$
$$
{\cal{E}}^i \equiv \frac{1}{2}\vep^{ijk}(B_{jk}+F_{jk}(c)) \eqno (21)
$$
Operators ${\cal{E}}^i$ and ${\cal{B}}^i$ obey the algebra of the electric 
and magnetic operators of Maxwell theory, as expected.

Now we proceed with the geometric representation appropiate to the 
formulation we are dealing with. The secondary first class constraints (17), 
(18) generate, as one could expect, the time independent gauge 
transformations that correspond to equations (9)-(11). Within the spirit of 
the L.R., we seek for non-local gauge invariant operators that replace the 
gauge dependent canonical ones $A_i$, $C_i$ and $B_{ij}$. An adecuate choice 
consists on the Wilson loop:
$$
W(\gamma )=\exp (i\oint_\gamma dy^iA_i(y)) \eqno (22)
$$
where $\gamma$ is a closed spatial path, and the operator:
$$
\om(\Sigma ,\Gamma )=\exp (i\oint_\Gamma dy^iC_i(y))
\exp (i\iint_\Sigma d\Sigma_k\vep^{ijk}B_{ij}) \eqno (23)
$$
which depends on the spatial open surface $\Sigma$ whose boundary is $\Gamma$. 
Taking an infinitesimal loop $\delta\gamma$ and surface $\delta\Sigma$ one 
has:
$$
W(\delta\gamma)={\bf 1} +i\delta\sigma^{ij}F_{ij}(A)+\theta (\delta\sigma^2)
\eqno (24)
$$
$$
\om (\delta\Sigma ,\delta\Gamma )={\bf 1}+i\delta\Sigma^{ij}(F_{ij}(c)+B_{ij})
+\theta (\delta\Sigma^2)\eqno (25)
$$
which shows that $W$ and $\Omega$ encode the local gauge invariant content of 
$A_i$, $C_i$ and $B_{ij}$, as claimed.

From the canonical commutators, it is easy to calculate the algebra obeyed by 
$W$ and $\Omega$:
$$
W(\gamma )\om (\Sigma ,\Gamma )=e^{i{\cal{L}}(\gamma ,\Gamma )}
\om (\Sigma ,\Gamma )w(\gamma )\eqno (26)
$$
where:
\setcounter{equation}{26}
\begin{eqnarray}
{\cal{L}}(\gamma ,\Gamma )&=&\oint_\gamma dx^i\iint_\Sigma d\Sigma_i(y)
\delta^3(x-y)\nonumber\\
&=&\frac{1}{4\pi}\oint_\gamma dx^i\oint_\Gamma dy^j\vep_{ijk}
\frac{(x-y)^k}{|x-y|^3}
\end{eqnarray}
is the linking number between the loops $\gamma$, $\Gamma$. Equation (26), 
which we shall call dual algebra (D.A.) of Maxwell theory, deserves the 
following comments. In virtue of constraints (17) one has that on the 
physical (i.e., gauge invariant) subspace:
$$
\om (\Sigma_{closed})|\Psi_{physical}>=|\Psi_{physical}>\eqno (28)
$$
that is to say, $\Omega$ does not depend on $\Sigma$, but only on its 
boundary $\Gamma$. Thus, $\Omega (\Gamma ,\Sigma )$, which by construction 
is the quantum operator associated to the electric flux through $\Sigma$, can 
also be viewed as the ``dual'' Wilson Loop, i.e., as the contour integral of 
the dual potential $\wt{A}$ along $\Gamma$. Observe, however, that the 
formulation does not include this potential as a lagrangian variable, which 
would be redundant.

It should be noticed the topological character of the D.A.. Both $W$ and 
$\Omega$, inasmuch as the linking number ${\cal{L}}(\gamma ,\Gamma )$ are 
metric independent objects. The latter is, in fact, a ``link invariant'', in 
knot theoretical terms. Of course, this does not mean that Maxwell theory is 
a topological one, since the Hamiltonian is constructed from the metric 
dependent combinations ${\cal{E}}^i{\cal{E}}^i$ and ${\cal{B}}^i{\cal{B}}^i$.

The algebra (26) may be realized by prescribing that $W$ and $\Omega$ act 
onto loop dependent wave functionals $\Psi (\gamma )$ as:
$$
W(\gamma_2)\Psi (\gamma_1)=\Psi (\gamma_2\cdot\gamma_1)\eqno (29)
$$
and
$$
\om (\Gamma )\Psi (\gamma_1)=e^{-i{\cal{L}}(\Gamma ,\gamma_1)}\Psi (\gamma_1)
\eqno (30)
$$

Here, $\Psi (\gamma )\equiv <\gamma |\Psi >$ is the wave functional in the 
usual L.R. $|\gamma >$ [12], while $\gamma_1 \cdot \gamma_2$ denotes the 
Abelian Group of Loops product [12]. To check that (29) (30) realize the 
D.A. (26) we have used the following elementary properties of the linking 
number:
$$
{\cal{L}}(\gamma_1,\gamma_2)={\cal{L}}(\gamma_2,\gamma_1)\eqno (31)
$$
$$
{\cal{L}}(\gamma_1,\gamma_2\cdot\gamma_3)={\cal{L}}(\gamma_1,\gamma_2)+
{\cal{L}}(\gamma_1,\gamma_3)\eqno (32)
$$
Equations (29), (30) have a simple geometrical meaning: $W(\gamma )$ performs 
a finite translation in Loop space, while $\Omega (\Gamma )$ (we suppress the 
spurious $\Sigma$-dependence) acts multiplicatively, by measuring how $\Gamma$ 
links with the wave functional argument. These roles are exchanged if one 
choose to work in the dual L.R. $|\Gamma >$: from equation (26), we see that 
the replacements $W(\gamma )\ole \om(\Gamma )$ and 
${\cal{L}}(\gamma , \Gamma )\to -{\cal{L}}(\gamma , \Gamma )$ leave the D.A. 
invariant. This is the way how self-duality of $D=4$ Maxwell theory is 
reflected in the present formulation.

It is worth mentioning that the D.A. (26) is closely related to the well 
known 't Hooft algebra, [10] which was obtained in the context of pure 
Yang-Mills theory. However, although the 't Hooft analogous to our 
$\Omega (\Gamma )$ (the $B$ operator in 't Hooft notation) can be interpreted 
as measuring in some sense the chromo-electric flux, it is not possible, as 
far as we know, to write it down in terms of the chromo-electric field. 
Hence, unlike the present case, $W$ and $B$ cannot describe the phase space, 
nor the Hamiltonian can be written in terms of them. This feature is related 
to the fact that there not exists a simple non-Abelian analogous to 
electric-magnetic duality [7,8].

To complete the description of quantum Maxwell theory within the above 
formulation, we have to express $H$ in terms of $W$ and $\Omega$. To this 
end, we recall the loop derivative of Gambini-Tr\'{\i}as [12]:
$$
\Delta_{ij}(x)f(\gamma )=\lim_{\sigma^{ij}\to 0}
\frac{f(\delta\gamma\cdot\gamma )-f(\gamma )}{\sigma^{ij}}\eqno (33)
$$
that measures how a loop-dependent object $f(\gamma )$ changes when $\gamma$ 
is modified by appending a small contour $\delta\gamma$ of area 
$\sigma^{ij}$ at $x$. Then the electric and magnetic field operators are 
given by:
$$
{\cal{B}}^k(x)=-i\vep^{ijk}\Delta_{ij}(x)w(\gamma )|_{\gamma =0}\eqno (34)
$$
$$
{\cal{E}}^k(x)=-i\vep^{ijk}\Delta_{ij}(x)\om(\gamma )|_{\Gamma =0}\eqno (35)
$$
and a loop-dependent Schr\"odinger equation can be written down, either in 
the $|\gamma >$ or the $|\Gamma >$ loop representations.

To conclude, let us briefly sketch how this ideas may be generalized to 
abelian gauge theories of arbitrary rank and dimension. To save some writing, 
we turn to the more compact form-notation. Consider, in $D=d+1$ dimensions, the 
theory associated to a $p$-form $A\ (p<D)$ whose lagrangian is given by:
$$
{\cal{L}}(A)=\alpha dA\wedge^*(dA)\eqno (36)
$$

As it is well known, the dual theory that correspond to (36) is described by 
the lagrangian ${\cal{L}}(\wt{A})$, where $\wt{A}$ is a $D-(p+2)$ form. It 
follows that for even $D$, self-dual theories are constructed from 
$\frac{D-2}{2}$ forms. The first order gauge-invariant lagrangian 
${\cal{L}}''$ (see equation (8)) associated to (36) is:
$$
{\cal{L}}''=\alpha dA\wedge B+\beta (B+dC)\wedge^*(B+dC)\eqno (37)
$$
where $B$ and $C$ are $D-(p+1)$ and $D-(p+2)$ forms respectively. $\cal{L}''$ 
is invariant under:
$$
\delta A=d\wedge \eqno (38.a)
$$
$$
\delta B=d\xi \eqno (38.b)
$$
$$
\delta C=\xi \eqno (38.c)
$$
As in Maxwell case, the dinamical equations obtained from $\cal{L}$ and 
$\cal{L}''$ are equivalent.

Canonical quantization follows along the same lines as before. Our main 
concern, which is to show the D.A. appropiate to the theories we are dealing 
with, may be resumed as follows. The first class constraints that arise tell 
us that the non-local operators carrying the gauge invariant content of the 
pull-backs of $A$, $B$ and $C$ to the spatial manifold ${\cal{R}}^d$, can be 
taken as:
$$
W(S_a)=\exp (i\int_{S_a}A) \eqno (39)
$$
$$
\om (S_b)=\exp (i\int_{S_b}(B+dC)) \eqno (40)
$$
where $S_a$ is a closed $p$-surface embeded in ${\cal{R}}^d$, while $S_b$ is an 
open $d-p$ one. As before, one of the constraints (the pull-back of $dB=0$) 
allows to show that on the physical subspace, $\om$ only depends on the 
boundary $\partial S_b$ of $S_b$. Then the D.A. may be written as:
$$
W(S_a)\om(\partial S_b)=e^{i{\cal{L}}(S_a,\partial S_b)}\om (\partial S_b)
W(S_a) \eqno (41)
$$
where:
$$
{\cal{L}}(S_a,\partial S_b)\equiv \frac{1}{k}\int_{S_a}d\Sigma^{i_1\cdots i_p}
(x)\int_{\partial S_b}d\Sigma^{i_{p+2}\cdots i_d}
\frac{(x-y)^{i_{p+1}}}{|x-y|^d}\vep_{i_1\cdots i_d} \eqno (42)
$$
measures the linking number of $S_a$ and $\partial S_b$ ($k$ is the area of 
the unit $d$-sphere). Equation (42) does not include the cases where either 
$A$, $B$, or both two are $0$-forms. To seek for the appropriate expressions 
for that cases, it must be observed that the boundary of a 1-dimensional 
``open surface'' (i.e., a open path) is a pair of oriented points $(x,y)$. 
Thus, when $A$ is a $0$-form $\varphi (x)$, its associated $W(S_a)$ will be:
$$
W((x,y))=\exp i\left \{\varphi (x)-\varphi (y) \right\} \eqno (43)
$$
and the correponding linking number for the case $A=\varphi (x)$, $d\geq 2$ 
is:
$$
{\cal{L}}((x,y),\partial S_b)=\frac{1}{k}\int_{\partial S_b}
d\Sigma_{(z)}^{i_1\cdots i_{d-1}}(z)\vep_{i_1\cdots i_d} 
\left \{\frac{(z-y)^{i_d}}{|x-y|^d}-\frac{(z-x)^{i_{d}}}{|x-y|^d}\right \} 
\eqno (44)
$$
This expression counts how many times the $d-1$ surface $\partial S_b$ 
encloses the points $x$ and $y$, taking into account that they contribute 
with opposite signs.

On the other hand, for $d=1$, where $A$ is forced to be a $0$-form, both $W$ 
and $\om$ result to be operators depending on pairs of points. The 
corresponding D.A. is given by:
$$
W((x,y))\om((x',y'))=\exp i\left \{ \theta (x-x')-\theta (x-y')-\theta (y-x')
+\theta (y-y') \right \} \eqno (45)
$$
The combination of Heaviside functions $\theta$ appearing in (45) is a 
topological invariant associated to the pairs $(x,y)$, $(x',y')$: it measures 
whether or not the line segment that corresponds to one of the pairs contains 
some point of the other pair. This quantity is the appropriate ``linking 
number'' for points on a line.

Summarizing, we have exploited the geometrical language of the L.R. to show 
that the ``electric magnetic'' duality of usual abelian gauge theories leads 
in a natural manner to a topological algebra of elementary observables that 
may replace the gauge-dependent canonical algebra. This algebra admits a 
realization in terms of operators acting on functionals that depend on 
geometrical objects, which is explicitly shown for the four dimensional Maxwell 
Theory.

The above ideas can be extended without mayor difficulties to the case of 
massive theories (i.e., Proca's model and its higher rank generalizations). 
It remains to be studied whether or not the present formulation can be useful 
to explore the analogous to electric-magnetic duality in non Abelian theories.

\end{document}